\documentclass{ws-procs10x7}

\newcommand{\eV}{\mbox{eV}}
\newcommand{\best}{{\mbox{\scriptsize best}}}
\newcommand{\GWthtyr}{\mbox{GW}_{\mbox{\scriptsize th}}\cdot\mbox{ton}\cdot\mbox{year}}
\newcommand{\sys}{{\mbox{\scriptsize sys}}}
\newcommand{\visi}{{\mbox{\scriptsize visi}}}
\newcommand{\sigS}{\sigma_{\mbox{\scriptsize S}}}
\newcommand{\sigBG}{\sigma_{\mbox{\scriptsize BG}}}
\newcommand{\sigdb}{\sigma_{\mbox{\scriptsize db}}}
\newcommand{\sigDb}{\sigma_{\mbox{\scriptsize Db}}}
\newcommand{\sigdB}{\sigma_{\mbox{\scriptsize dB}}}
\newcommand{\sigDB}{\sigma_{\mbox{\scriptsize DB}}}
\newcommand{\sigD}{\sigma_{\mbox{\scriptsize D}}}
\newcommand{\sigB}{\sigma_{\mbox{\scriptsize B}}}
\newcommand{\sigd}{\sigma_{\mbox{\scriptsize d}}}
\newcommand{\sigb}{\sigma_{\mbox{\scriptsize b}}}

\begin{document}

\title{
 Exploring Leptonic CP Violation
with Combined Analysis of Reactor and Neutrino Superbeam Experiments
%Exploring Leptonic CP Violation
%by Reactor and Neutrino Superbeam Experiments
}

\author{Hiroaki Sugiyama}

\address{Theory Group, KEK, Tsukuba, Ibaraki 305-0801, Japan\\
E-mail: hiroaki at post.kek.jp}

\twocolumn[\maketitle\abstract{
 We investigate the possibility to find the leptonic CP-violation
by combining the reactor experiment with the superbeam experiment
without antineutrino superbeam.
 We show also how much the sensitivity on CP-violating phase $\delta$
is affected by the fact that we have not known
the sign of $\Delta m^2_{31}$.}]

\section{Introduction}

 Observing the leptonic CP-violation is
one of aims in future experiments of the neutrino oscillation.
 It can be achieved by comparing oscillation probabilities
with neutrino beam and its antiparticle one;
 For example, T2K experiment will measure
$P(\nu_\mu\to\nu_e)$ and $P(\overline{\nu}_\mu\to\overline{\nu}_e)$
precisely in its phase II\@.
 Such a comparison is important obviously
because it gives a direct observation of the leptonic
CP-violation (except the matter effect mimics the CP-violation).
 It seems, however, the matter of the simple method
that the exposure with antineutrino beam needs to be
about three times longer than that with neutrino beam
because of its smaller detection cross-section;
 In phase II of T2K experiment,
about 6year exposure with $\overline{\nu}_\mu$ is
planed after 2year exposure with $\nu_\mu$ beam.
 The smaller cross-section even gives worth S/N ratio.
 Thus, it seems fruitful to consider other possibilities
to see CP-violation.

 In the scheme of three neutrino oscillation,
the CP-violation is controlled by the CP-violating phase $\delta$.
 If we can know $\delta$ is not vanishing,
it means a measurement of CP-violation indirectly
by assuming three neutrino oscillation.
 The information about $\delta$ is included in $P(\nu_\mu\to\nu_e)$,
but the oscillation probability include $\theta_{13}$ also
as the parameter to be determined.
 That is why we require additionally the measurement of
$P(\overline{\nu}_\mu\to\overline{\nu}_e)$,
whose parameters to be determined are also $\delta$ and $\theta_{13}$,
to extract the value of $\delta$.
 Any oscillation probability, however, can be the additional one
as long as it has information on the value of $\theta_{13}$.
 In this talk\cite{Minakata:2003wq},
we consider a combination of a superbeam experiment
with neutrino beam and a reactor experiment,
which is a pure measurement of $\theta_{13}$,
to explore the CP-violation in lepton sector.
 Note that the method have an advantage of speed
because the reactor experiment can run parallel to
the superbeam $\nu$ experiment
in contrast with the superbeam $\overline{\nu}$ experiment.
 Although we can extract the value of $\delta$
from the combination of reactor and superbeam $\nu$ experiments
in principle, a quantitative analysis is necessary for concreteness.

\begin{table*}
\begin{tabular}{|c|c|c|c||c|}
 \hline
 \multicolumn{2}{|c|}{} & \multicolumn{2}{|c||}{between detectors} & \\
 \cline{3-4}
 \multicolumn{2}{|c|}{} & correlated & uncorrelated & \ single detector \ \\
 \hline
 \ between bins \ & correlated & \ $\sigDB = 2.5\%$ \ & \
 $\sigdB = 0.5\%$ \ & $\sigB \simeq 2.6\%$ \\
 \cline{2-5}
 & \ uncorrelated \ & $\sigDb = 2.5\%$ &
 $\sigdb = 0.5\%$ & $\sigb \simeq 2.6\%$ \\
 \hline\hline
 \multicolumn{2}{|c|}{ \ total number of events \ } &
 $\sigD \simeq 2.6\%$ &
 $\sigd \simeq 0.5\%$ &
 \ $ \sigma_\sys \simeq 2.7\%$ \ \\
 \hline
\end{tabular}
\caption{Listed are assumed values of systematic errors
$\sigDB$, $\sigDb$, $\sigdB$, and $\sigdb$.
 The subscripts D (d) and B (b) are represent
the correlated (uncorrelated) error among detectors and bins,
respectively.
 Using those four values,
the errors for the total number of events and for single detector
are calculated. (See Appendix.)
 $\sigd \simeq 0.5\%$ means 0.8\% relative normalization error
on the comparison of numbers of events at near and far detectors:
$0.8\% \simeq \sqrt{2}\,\sigd$.
 $\sigma_\sys = 2.7\%$ corresponds to the systematic error
in CHOOZ experiment.}
\label{syserror}
\end{table*}

\section{Settings}

 As a superbeam experiment,
we deal with phase II of T2K experiment
without $\overline{\nu}_\mu$ beam;
 The beam power is assumed to be 4MW,
the fiducial volume of detector (Hyper-Kamiokande) is 540kt,
and the exposure time is 2 years with off-axis 2deg.\ $\nu_\mu$ beam.
 Total number of events within 0.4-1.2GeV is used for the analysis,
and we assume 2\% systematic errors for estimations
of numbers of signal and background events:
$\sigS = \sigBG = 2\%$. 
 Throughout this talk,
oscillation parameter values are fixed as follows:
$|\Delta m^2_{31}| = 2.5\times 10^{-3}\eV^2$,
$\Delta m^2_{21} = 7.3\times 10^{-5}\eV^2$,
$\tan^2{\theta_{12}} = 0.38$ $(32^\circ)$,
$\sin^2{2\theta_{23}} = 1$ $(90^\circ)$.
 Earth matter density is chosen as $\rho = 2.3 \mbox{g}\cdot\mbox{cm}^{-1}$.

 On the other hand,
we deal with a simple complex of one reactor and two detectors
as a future reactor experiment.
 The simple set-up is a good approximation
if the set-up of the future reactor experiment is appropriate enough.
 Since the determination of $\delta$ requires
very precise measurement,
the position of the far detector is assumed to be
optimal one which is 1.7km away from the reactor
and the scale of the reactor experiment is assumed
to be rather large, $\sim 10^3\GWthtyr$
('thermal power of the reactor' times 'detector volume'
times 'exposure time').
 $\overline{\nu}_e$ detection efficiency is assumed to be 70\%.
 Furthermore,
we rely upon spectral information also
for the precise determination of $\theta_{13}$.
 We use 14 bins of 0.5MeV width in 1-8MeV visible energy:
$E_\visi = E_{\overline{\nu}_e} - 0.8\mbox{MeV}$.
 For the analysis,
four types of systematic errors
($\sigDB, \sigDb, \sigdB, \sigdb$) should be considered at least.
 An example of $\sigDB$ is the error in reactor power,
which gives a common effect on numbers of events in all bins
at each detector.
 $\sigDb$ is, for example, the error in energy dependence (shape)
of flux or cross-section, which is bin-by-bin uncorrelated
but correlated between detectors.
 A typical origin of $\sigdB$ is the error in detector volume,
which has overall effect for all bins
but is uncorrelated between detectors.
 $\sigdb$ is completely-uncorrelated error
and somewhat accidental one;
 Such a error dominates the sensitivity
because it can not be cancelled by any comparison (detectors, bins).
 Assumed values of those errors are listed in Table~\ref{syserror}.
 Note that if $\sigdb$ is set to be zero,
the sensitivity does not saturate even for extremely long exposure
and then unrealistically high sensitivity is obtained.

\section{Reactor-superbeam combined analysis}

\begin{figure}%1
\begin{center}
\epsfxsize190pt
\figurebox{120pt}{160pt}{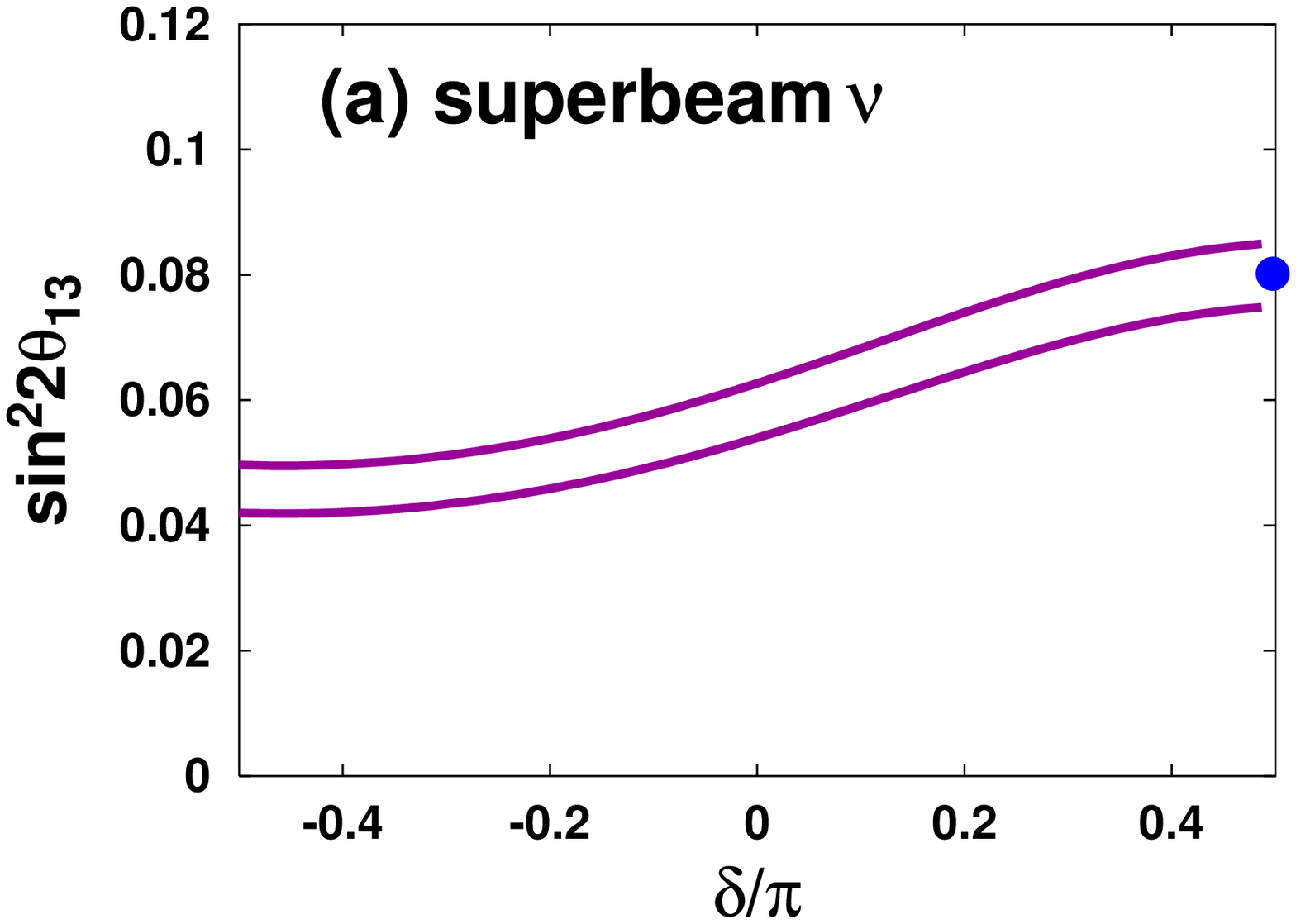}\hspace*{0mm}\\
\epsfxsize190pt
\figurebox{120pt}{160pt}{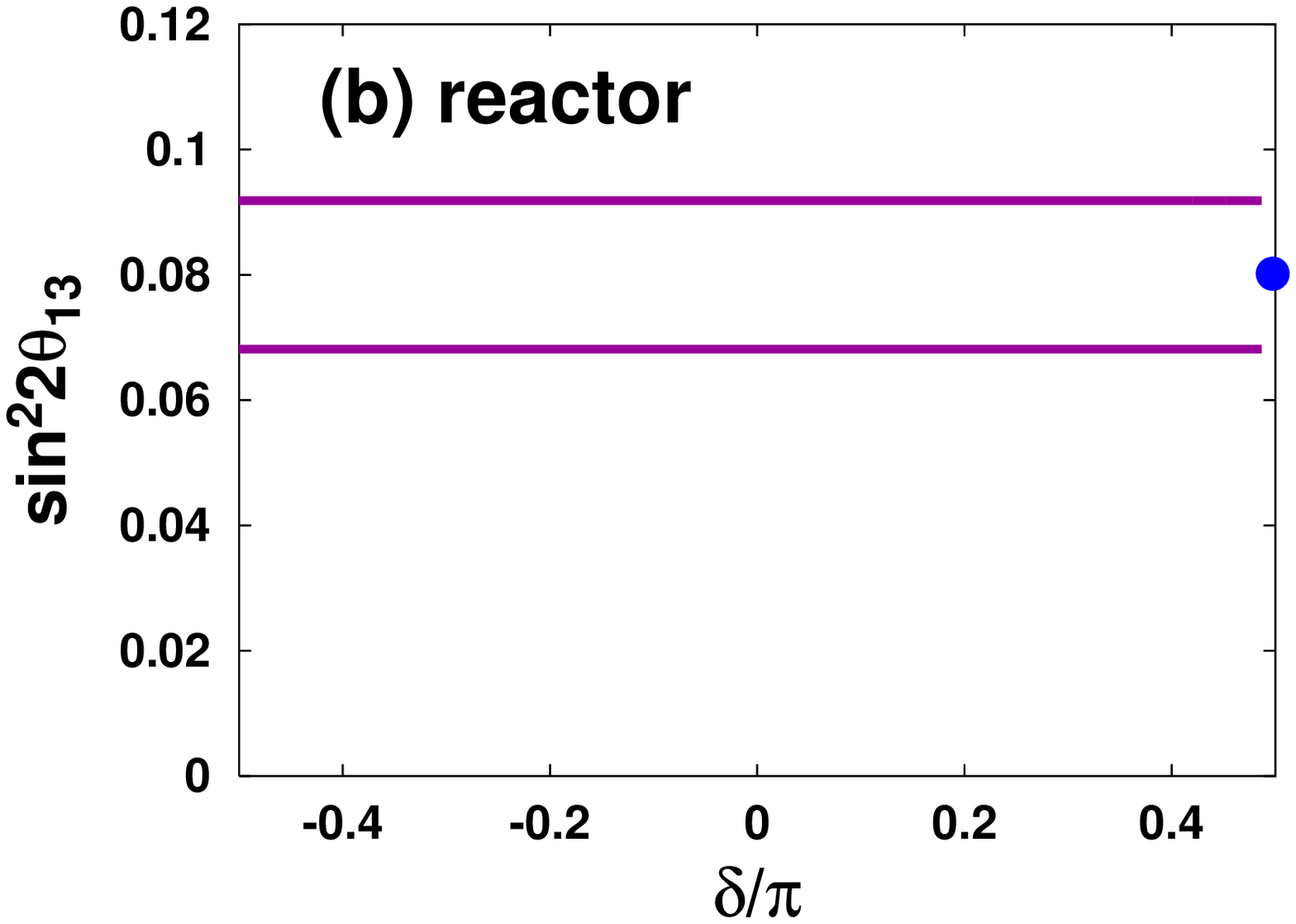}\hspace*{0mm}\\
\epsfxsize190pt
\figurebox{120pt}{160pt}{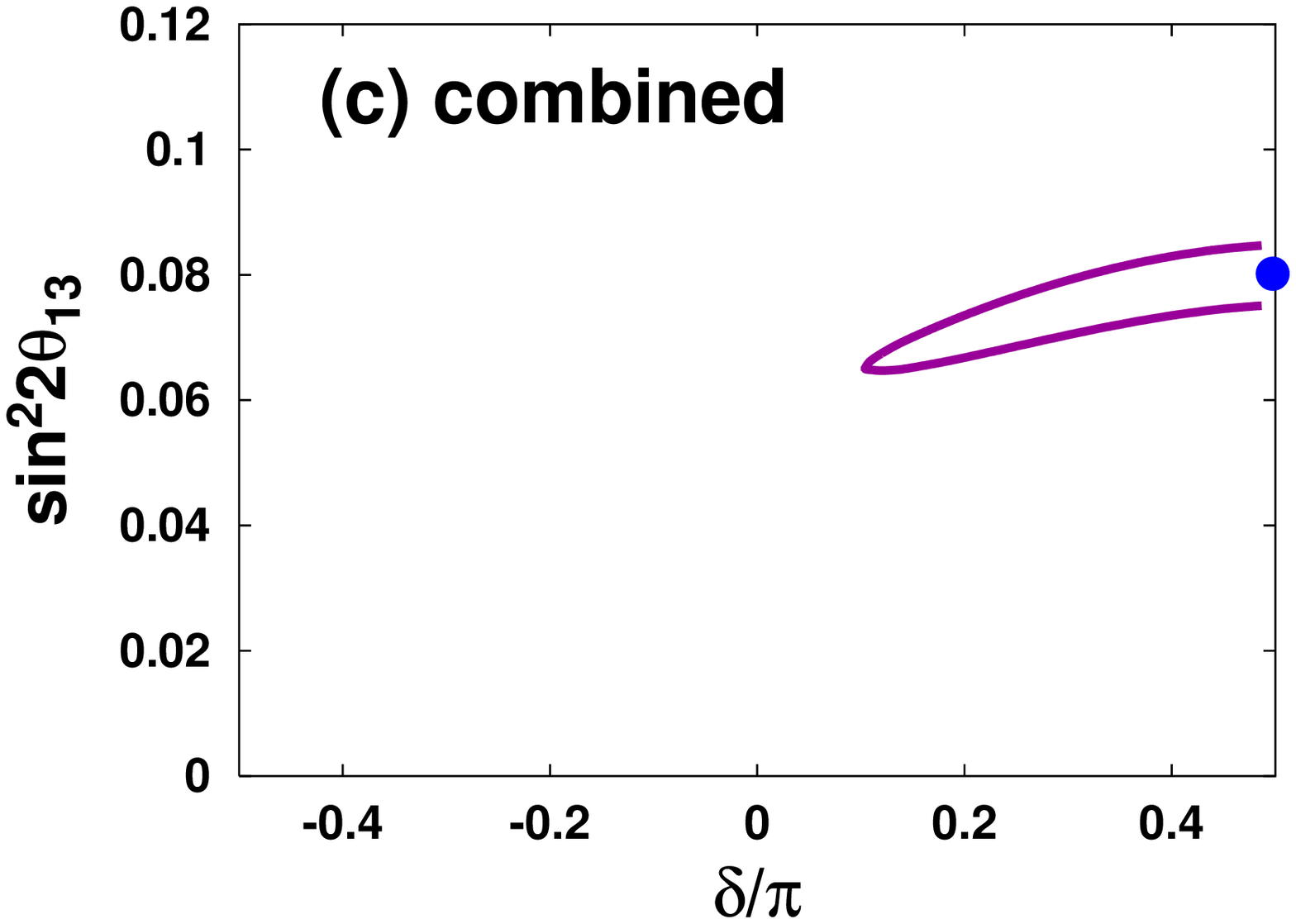}\hspace*{0mm}
\epsfxsize190pt
\end{center}
\caption{ The true values are assumed to be
$\sin^2{2\theta_{13}^\best} = 0.08$ and $\delta^\best = \pi/2$
as an example.
(a) shows the allowed region for a given $P(\nu_\mu\to\nu_e)$
to be obtained in the superbeam experiment.
 The allowed region to be obtained by reactor experiment is presented
in (b).
 (c) is the result of the combined analysis. 
}
\label{fig:comb}
\end{figure}

 Fig.~\ref{fig:comb} shows how the combined analysis works;
 Best-fit values of $\theta_{13}$ and $\delta$,
which are chosen by nature, are assumed to be
$\sin^2{2\theta_{13}^\best}=0.08$ and $\delta^\best=\pi/2$
in Fig.~\ref{fig:comb}, respectively.
 Fig.~\ref{fig:comb}(a) shows an allowed region
that we obtain when $P(\nu_\mu\to\nu_e)$ is measured
in T2K phase II\@.
 The allowed region depends on $\delta$ through
Jarlskog factor.
 On the other hand, the reactor experiment
gives another allowed region
as is shown in Fig.~\ref{fig:comb}(b). 
 The allowed region is independent of $\delta$
because the reactor experiment is a pure measurement
of $\theta_{13}$.
 Roughly speaking,
the overlap between those two allowed regions
results in the allowed region obtained by the combined analysis.
 Fig.~\ref{fig:comb}(c) shows the actual result obtained
by the combined analysis for the input values of
$\theta_{13}$ and $\delta$.
 Since $\delta = 0$ is excluded
by the combined analysis in this case,
we find that CP is violating in the lepton sector.
 Then, we want to know which values of $\theta_{13}^\best$
and $\delta^\best$ exclude the hypothesis $\delta = 0$
by the combined analysis.

\begin{figure}
\begin{center}
\epsfxsize190pt
\figurebox{120pt}{160pt}{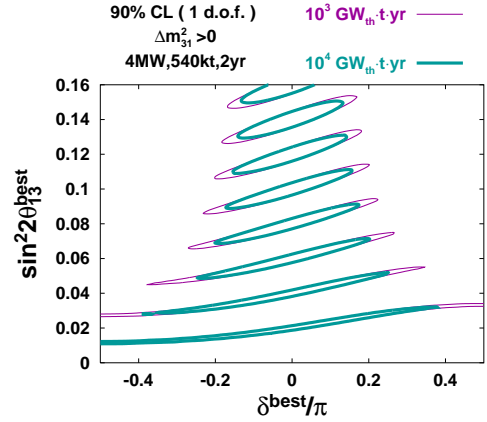}\hspace*{0mm}\\
\epsfxsize190pt
\end{center}
\caption{Shown are the regions consistent with the hypothesis
$\delta = 0$ at 90\%CL by the reactor-superbeam combined analysis.
 Thin and bold lines are for $10^3$ and $10^4\GWthtyr$ exposure
of the reactor experiment.
If nature chooses outside of those regions, we know CP is violating.}
\label{fig:rCP1}
\end{figure}

 Fig.~\ref{fig:rCP1} shows the regions that are consistent
with the hypothesis $\delta = 0$ at 90\%CL\@.
 Therefore, we find that CP is violating,
if nature chooses the values of $\theta_{13}^\best$
and $\delta^\best$ outside of the envelope of those regions.
 If $\delta^\best$ is very close to $\delta = 0$,
we can not distinguish them.
 If $\theta_{13}^\best$ is too small,
$\delta$ is not so much restricted by the combined analysis
because of small Jarlskog factor, namely small $\delta$-dependence
of the allowed region obtained by T2K phase II with $\nu_\mu$ beam.
 Fig.~\ref{fig:rCP1} is consistent with those qualitative expectation.
 We see in Fig.~\ref{fig:rCP1} that
we can find leptonic CP-violation at 90\%CL
if $\sin^2{2\theta_{13}^\best} \ge 0.05$ $(6^\circ)$ and
$\delta^\best \ge 0.3\pi$ $(54^\circ)$.

\begin{figure}
\begin{center}
\epsfxsize190pt
\figurebox{120pt}{160pt}{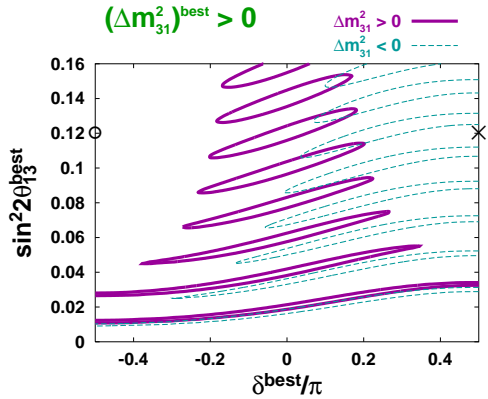}\hspace*{0mm}\\
\epsfxsize190pt
\end{center}
\caption{
 The contours are regions consistent with $\delta = 0$.
 The true sign of $\Delta m^2_{31}$ is assumed as
$(\Delta m^2_{31})^\best > 0$.
 Solid lines are for fitting with $\Delta m^2_{31} > 0$,
and dashed-lines are for fitting with $\Delta m^2_{31} < 0$.
}
\label{fig:rCP2}
\end{figure}

 Actually,
we fixed the sign of $\Delta m^2_{31}$ as positive
in Fig.~\ref{fig:rCP1}.
 The sign has, however, not been determined yet.
 Thus, we must use each sign of $\Delta m^2_{31}$
for fitting even if nature chooses positive value
because we do not know the nature's choice.
 Fig.~\ref{fig:rCP2} shows the result for
the case of unknown sign of $\Delta m^2_{31}$.
 If the sign of $\Delta m^2_{31}$ is known,
we can use solid lines only.
 Then, for example, we find CP is violating
if nature chooses the values indicated by circle or cross
in Fig.\ref{fig:rCP2}.
 For the case of unknown sign of $\Delta m^2_{31}$,
we must use dashed lines also, and the point of cross mark
enters their region.
 It means that even though nature chooses $\delta^\best = \pi/2$
(maximal CP-violation), it can be explained by $\delta = 0$
with wrong sign of $\Delta m^2_{31}$.
 Roughly speaking, almost a half of sensitivity region
is contaminated by the fitting with wrong sign of $\Delta m^2_{31}$.
 Since it is caused by $P(\nu_\mu\to\nu_e)$,
the contamination occurs even for the conventional method
with antineutrino superbeam.

\section{Conclusions}

 In this talk, we considered determining leptonic CP-violation
by using combined analysis of reactor and superbeam experiments
with neutrino beam.
 We found that it is possible with the method to know that
CP-violating phase $\delta$ is not vanishing at 90\%CL
if nature chooses $\sin^2{2\theta_{13}^\best} \ge 0.05$
and $\delta^\best \ge 0.3\pi$ $(54^\circ)$ as the true values.
 Actually,
the sensitivity is worse than that of conventional method
(99.73\%CL determination for
$\sin^2{2\theta_{13}^\best} \ge 0.02$ and $\delta^\best \ge 20^\circ$)
in which superbeam $\nu$ experiment is combined with
subsequent long-term experiment of $\overline{\nu}$ superbeam.
 The reactor-superbeam combined method, however, can give
earlier information on $\delta$ because the reactor experiment
can start before the finish of superbeam $\nu$ experiment.
 Such a information will be helpful
for later precise measurement of $\delta$ with conventional method.
 Therefore, the new combined method seems to be worth doing.

 We should keep in our mind that unknown sign of $\Delta m^2_{31}$
makes the sensitivity on $\delta$ worse very much
even with rather short baseline experiment such as T2K (295km).
 It is the problem not only for reactor-superbeam combined method
but also for conventional method.

\section*{Acknowledgments}
 This talk was supported by the Research Fellowship
of Japan Society for the Promotion of Science (JSPS)
for Young Scientists.

\begin{appendix}
%\section*{Appendix}
 The following is how to calculate $\sigB$, $\sigb$,
$\sigD$, $\sigd$, and $\sigma_\sys$
from $\sigDB$, $\sigDb$, $\sigdB$, and $\sigdb$.
\begin{eqnarray*}
&&
\sigB^2 = \sigDB^2 + \sigdB^2, \ \
\sigb^2 = \sigDb^2 + \sigdb^2,\\[2mm]
&&
\sigD^2 = \sigDB^2 + \sigDb^2 \frac{ \sum_i (N_{ni}^\best)^2 }
                                   { \left( \sum_i N_{ni}^\best \right)^2 },\\[2mm]
&&
\sigd^2 = \sigdB^2 + \sigdb^2 \frac{ \sum_i (N_{ni}^\best)^2 }
                                   { \left( \sum_i N_{ni}^\best \right)^2 },\\[2mm]
&&
\sigma_\sys^2 = \sigD^2 + \sigd^2
 = \sigB^2 + \sigb^2 \frac{ \sum_i (N_{ni}^\best)^2 }
                            { \left( \sum_i N_{ni}^\best \right)^2 }.
\end{eqnarray*}
 $N_{ni}^\best$ denotes the number of signal events within $i$th bin
at near detector, which calculated for best-fit (input) values
of parameters as an ``experimental data''.
 In our analysis, the coefficient of $\sigDb^2$ is about 1/9.
\end{appendix}

\end{document}